# The search process of professional detection dog teams in realistic environments

Stephan P. Kaufhold, Jack Terwilliger, Samantha Kefer, Federico Rossano

Department of Cognitive Science, University of California San Diego, California, USA

**Abstract**

Scent detection work is a form of interspecies cooperation. Dogs sample odors that handlers cannot perceive, and handlers decide where the team searches and which of the dogs' responses counts as a find. However, detection performance is predominantly measured by outcomes and dog characteristics. To better observe the search process itself, we collected data at two judged detection events (109 searches of 37 professional narcotics teams). All searches were video-recorded and included an outdoor area, a vehicle, and a building search, each with two targets and food and currency distractors. We synchronized computer-vision tracking of dogs and handlers with continuous behavioral coding. We observed variation across teams and in total 42% of targets were found. Finds were rarely missed because dogs never came close to a target or because handlers did not call an indication. Rather, most were missed at the encounter itself, as dogs came within 1 m of nearly every target but indicated on only about one in six of these passes. Handlers did not behave differently near targets than near distractors before their dogs responded. Finding targets was moderately consistent within teams, whereas false calls depended on the individual search and on specific items. Observing the search process suggests that success is a result of the dog's response, the handler's decision, and their coordination.

## Introduction

Perception and cognition evolved to guide animals' behavior. Each species perceives the environment in relation to its own sensory, motoric, and cognitive capabilities, i.e., its distinct umwelt (von Uexküll, 1934/2010). Therefore, the same physical environment affords different actions to different organisms (Gibson, 1979). When individuals from different species cooperate, their different perceptual, behavioral, and cognitive capabilities can complement each other. This can expand the shared affordances beyond what either individual could achieve alone. This is often the case for collaboration between dogs and humans and likely explains the shared evolutionary history of both lineages (e.g., Chambers et al., 2020; Freedman & Wayne, 2017). For example, both species differ markedly in their sensory capabilities, especially in olfaction (Craven et al., 2010). However, domestication led dogs to attend to human communicative signals more readily, and earlier in development, than wolves, their closest wild relatives (Hare & Tomasello, 2005; Miklósi & Topál, 2013; Salomons et al., 2021). Humans have built working relationships around dogs' specific abilities, and scent detection has become one of the most prominent roles of modern working dogs, with applications in law enforcement, military operations, medical detection, and conservation (Beebe et al., 2016; Hayes et al., 2018; Lazarowski et al., 2020a). Within scent detection work, the dog samples an odor environment that the handler cannot perceive. In turn, the handler decides where the dyad goes, when a location has been sufficiently covered, and further must interpret whether and when the dog's behavior counts as a *find*. Therefore, scent detection can be understood as a form of interspecies coordination where the two partners perceive and act on the same search space from different, only partly overlapping, sensory and cognitive perspectives with different responsibilities.

The division of labor between handler and dog can be described in terms of distributed cognition. According to this view, a task is accomplished by a system of agents and objects rather than by a single mind (Hutchins, 1995). Amon and Favela (2019) put forth three criteria for treating a human-dog dyad as one cognitive system: (1) interaction-dominant dynamics (i.e., each partner's behavior continuously constrains the other's), (2) agency on both sides, and a (3) shared orientation to the task. For example, in guide-dog dyads the role of initiating an action passes back and forth between dog and handler many times over the course of a walk (Naderi et al., 2001). Within a detection search, the same target substance might afford different actions. For example, the dog can approach, sample and indicate on the target's odor plume, whereas the handler can present, detail, and call its location. The specific path of a search therefore emerges from the interaction of the handler's knowledge (e.g., where narcotics are commonly concealed in a vehicle) and the dog's perception (e.g., the direction from which odor arrives). The outcome of a search thus depends on the interaction as much as on either partner. Explaining why searches succeed thus requires observing where in a search the contributions of both handler and dog meet.

Most research on detection dogs has focused on individual differences, relating characteristics that are fixed before the search begins (e.g., breed, temperament, cognitive test scores, training history, handler experience) to summary outcomes (e.g., the proportion of targets found, a pass on a certification test; for reviews, see Brady et al., 2018; Bray et al., 2021; La Toya et al., 2017; Lazarowski et al., 2020b; Troisi et al., 2019). Breed effects have been documented in large samples of police dogs (Jezierski et al., 2014). However, a laboratory discrimination task yielded unexpected results with pugs significantly outperforming German shepherds (Hall et al., 2015). Other studies did not find significant breed differences in certified narcotics teams (Rice & Velasco, 2023). Several studies linked traits to detection dog performance, for example, inhibitory control, reward engagement, emotional stability, and independence from the handler (e.g., Lazarowski et al., 2020b, 2023; Roulaux et al., 2026; Tiira et al., 2020; Zheng et al., 2026). While such studies can provide indicators of which teams tend to succeed, their predictors are assessments outside of the search and thus reveal little about *how* a search is accomplished, i.e., its proximate mechanisms of success.

Another feature of this literature is that performance is often measured near its ceiling. For example, certified narcotics teams tested under familiar, single-blind conditions (i.e., the handler did not know the target locations, but the evaluator did) alerted correctly on over 90% of trials with about 10% false alerts (Rice & Velasco, 2021, 2023), and fully trained police dogs found hidden drugs in 87.7% of 1,219 searches (Jezierski et al., 2014). Performance within such studies confirm competence. However, they leave little variance to explain differences in strategies and teams. Testing conditions that more closely resemble real-world searches show a decline in performance. For example, teams that passed 94-100% of single-blind certification trials passed 72-88% under double-blind conditions (Quigley-McBride et al., 2026). Another study showed that no explosives team (N = 56) in a multi-site field study met the criterion of 90% detection with 10% false alerts (Karpinsky et al., 2025). Further, police dogs that were 83% correct in rooms were 58 to 64% correct at cars (Jezierski et al., 2014), and volunteer conservation teams lost about 20% sensitivity in each unfamiliar context (Rutter et al., 2021). Differences between teams are thus exacerbated under realistic conditions, and the search process, including the interaction between the two partners, should be studied under such conditions. Here, lower accuracy may reflect the demands of the situation as much as the abilities of the teams.

The literature treats the handler with suspicion but has rarely observed what handlers do during a search. Since Pfungst's (1911) analysis of Clever Hans, unintentional human cueing has been the standard explanation for animal performance that tracks human expectation. Lit et al. (2011) had 18 certified detection teams search rooms that contained no target odor; handlers reported 225 alerts across 144 runs, clustered at locations the handlers had been told were marked. The authors noted that their data could not separate two possible explanations, i.e., that handlers called alerts that their dogs never gave, or that handler belief cued their dogs to genuinely alert. These two accounts cannot be distinguished without video of the search. Methodological reviews therefore called for double-blind testing (Johnen et al., 2017; Lazarowski et al., 2020a). It has been shown that dogs were more likely to correctly alert when judges during certifications knew the correct hiding spots of target odors compared to searches where both judges and handlers were naïve to them (Quigley-McBride et al., 2026).

Both the dog and the handler play an active role in producing alerts during searches. When dogs detect a target odor, they produce a trained response (e.g., sitting, staring) and other behavioral changes, for example in their sniffing behavior (Concha et al., 2014) or tail movements (Martvel et al., 2025). For the dog's odor response to turn into an alert, the handler then needs to recognize and declare it. Rather than being a passive recipient of the dog's indications, the handler is therefore also an active detector within the team. It has been demonstrated that searches can be affected by both the handler's knowledge and state. For example, telling handlers how many hides were present changed how sport dogs searched without changing false alerts (DeChant et al., 2020). Another study showed that dogs of stressed handlers found explosives faster (Zubedat et al., 2014). Notably, both effects were inferred from outcomes under manipulated conditions.

Another limitation in the current literature lies in what is being observed and quantified. Most field and certification studies, including those cited above, score the search by the alert that the handler calls, which is usually recorded on a score sheet by an evaluator (Karpinsky et al., 2025; Lit et al., 2011; Quigley-McBride et al., 2026; Rice & Velasco, 2023). However, the dog's own indication is usually not recorded as an event of its own. Subsequent scoring from the sheet therefore does not distinguish the outcome (handler call) from the behavior that produced it. With video-recordings, "the dog alerted" can be broken down into several observable events, including where the dog was, its behavior, how long it took the handler to call, and how the call relates to the items around the dog. The search process itself has overall received less attention, with some exceptions. For example, Martin et al. (2022) tracked dogs from video recordings and related the distance and frequency of their search movements to their handlers' ratings of search efficiency. Jinn et al. (2020) and Mejia et al. (2024) showed that both ambient conditions and the physical properties of the odorant shape how dogs move. Jezierski et al. (2014) counted how often dogs passed within 1 m of a hide before indicating. Rating instruments for search behavior exist but are of limited reliability, because latent constructs such as *independence* or *control* are not reliably rated (Rooney & Clark, 2021) and handlers may rate their own dogs leniently (Clark et al., 2020). None of these studies observed professional teams in realistic environments while continuously coding the communication between handler and dog and tracking their movements in metric space.

Notably, observing the search process under field conditions is a trade-off. Here experimental control is reduced, but it increases ecological validity, which is best understood as a spectrum with several interacting dimensions (how representative the subjects are, how natural the environment is, and how natural the task and its interface are) rather than a dichotomy between laboratory and field (Kaufhold et al., 2026). Our goal for the present study was to have high ecological validity on all three dimensions while establishing conditions that make detailed measurements possible. Our subjects were professional detection teams working with their own dogs, procedures, and leash choices. The search environments were cluttered and contained attractive distractors (food items), and the task reflected the team's ordinary search procedure under a time limit.

Here, we report on our analysis of the search process and outcomes of two detection events, held one year apart, consisting of 37 professional narcotics teams and 109 searches. We reconstructed every search as a synchronized record of the handler's and the dog's positions in metric ground coordinates, obtained from fixed cameras with computer vision, and of their communicative behavior, coded continuously from video, and every handler alert (hereafter, call) was located and reviewed on video independently of the judges. We try to identify the relative contributions of handlers and dogs and their interactions by asking four questions: (1) How well do professional teams perform in realistic environments? (2) Where in the chain from the dog approaching a target, to the dog indicating, to the handler calling are finds missed? (3) Do handlers behave differently near targets than near distractors before the dog responds, as unintentional cueing would require, and do they call the indications their dogs give? (4) Which performance aspects are properties of the team rather than of the specific search, and which characteristics of dog and handler and which features of the search process predict them?

## Method

### *Ethics statement*

We obtained approval from the University of California, San Diego Institutional Review Board (805704) for all procedures involving human participants. In addition, all procedures involving dogs were approved by the University of California, San Diego Institutional Animal Care and Use Committee (S23142). Prior to the study, we obtained written informed consent for the study participation and audio- and video-recordings of all handlers. Throughout our data collection, dogs were handled exclusively by their own handlers, and search tasks were similar to the regular work and training of the teams.

***Participants***

We collected data at two separate detection events, which were held on a university campus in California in spring 2025 and spring 2026. Both events were organized by our laboratory in collaboration with a professional detection dog training company. We ran each event as a judged competition that was free to attend for professional detection dog handlers. The best teams, determined by scores assigned by judges, received trophies and small prizes. Handlers completed a paper questionnaire on their own experience and on their dog prior to any search, which included breed, age, sex, reproductive status, years of scent training, trained target odors, reinforcement, typical search terrain, frequency of on- and off-leash searches, certifications, and prior competitions.

Our study participants were a sample of North American professional detection dog handlers from law enforcement, corrections, and private security. The 2025 teams were all from law enforcement and corrections agencies; the 2026 teams were a mix of law enforcement and private security. To keep the two events comparable, we restricted the analyses to narcotics detection teams, the only target category present in both years. Teams that searched for other target categories (firearms, explosives) also took part but were not included in the current analysis due to their small numbers. Two handlers took part in both events with the same dog (dog age and scent-work experience one year further along on the second questionnaire). To keep each dyad's outcomes independent across events, we included only their 2026 searches in our analyses; however, including their 2025 searches changed no conclusion. Our analyses thus cover 37 teams consisting of 10 narcotics teams in 2025 (28 searches; one team completed only the outdoor search) and 27 in 2026 (81 searches; see Tables S1 and S2 for participant details). On average, handlers in 2026 had more years of detection-dog experience than handlers in 2025 (Mdn = 5.0 vs 2.0 years), and within 2026 the security handlers were more experienced (Mdn = 10.0 years) than the law enforcement handlers (Mdn = 3.0 years). In both years, most dogs were Belgian Malinois, followed by other shepherd breeds, and most were intact males working for a toy reward.

***Search environments and hides***

Every dog-handler team searched three environments: an outdoor area, an outdoor area with three vehicles, and an indoor room (Table 1; Figure 1A). Teams lined up outside a search area shortly before their turn. All search areas were occluded from view using fences or drapes, ensuring that no handler could observe a search area or witness another team's search before their own.

The searches took place in high-ceilinged event rooms on campus (different rooms in 2025 and 2026) and on flat asphalt parking lots. Neither event day had rain. The 2025 event took place on a sunny day and the 2026 event on a cloudy day, both with moderate temperatures. Both outdoor and indoor areas were populated with everyday objects, serving both as potential hiding places and as clutter (cardboard boxes, tin cans, chairs, bicycles, cabinets, drawers, tables, toolboxes, office equipment, etc.). The vehicle area contained three vehicles (in 2026 a cargo van, a compact van, and a minivan; in 2025 the same cargo van and two compact vans; models in Supplementary Methods S1.0), and handlers were informed that the cargo van was the only vehicle to be searched inside. In 2025, the cargo van was closed and handlers opened its side or rear doors, whereas in 2026 its rear doors were already open.

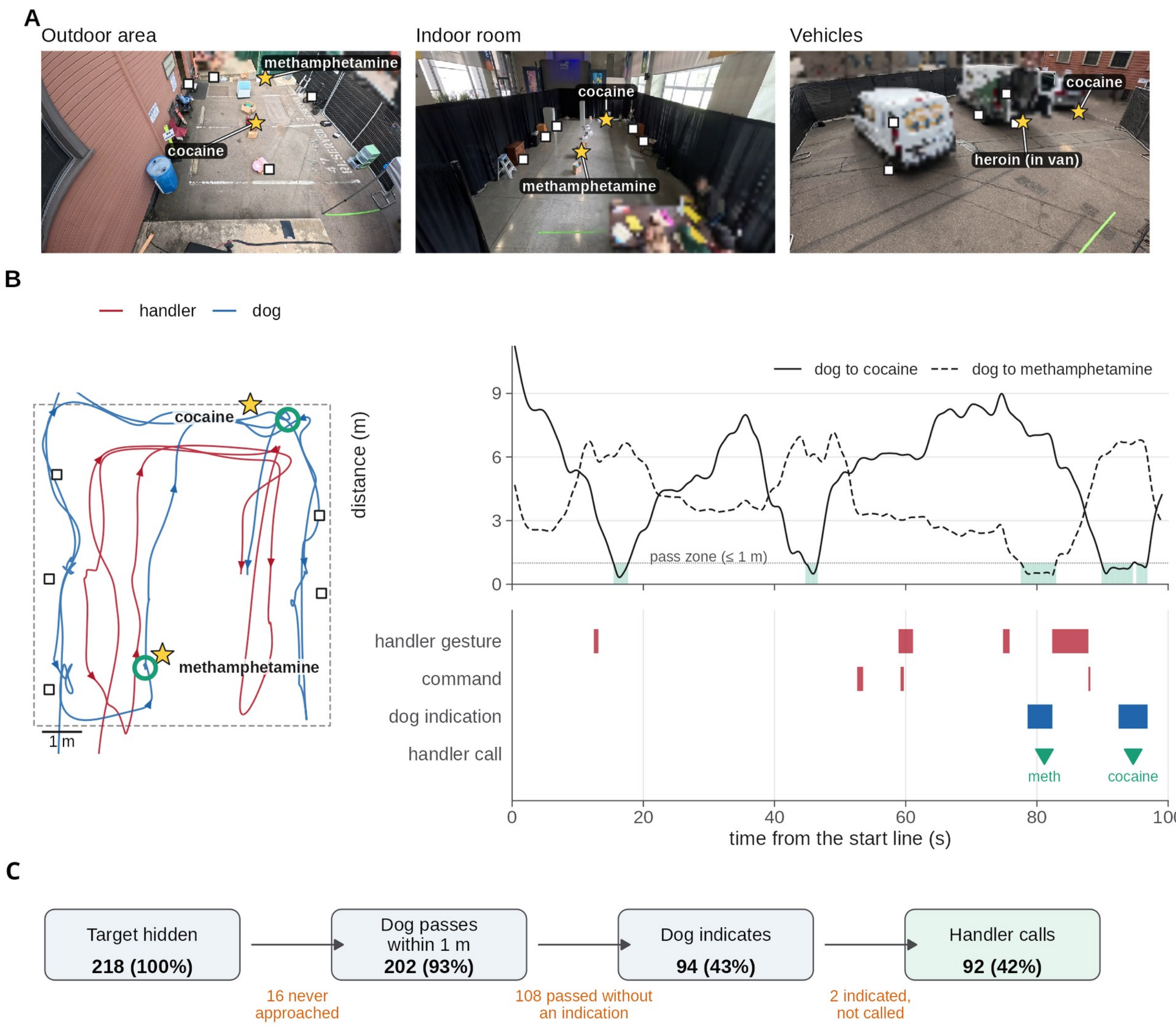

**Figure 1.** Study overview. (A) The three 2026 search environments seen from one of the fixed cameras with the targets (stars) and distractors (squares) of that search (items inside the cargo van are marked at the van). (B) A 2026 indoor search (Supplementary Video S3) as synchronized data: the tracked paths of dog (blue) and handler (red) on the floor plan, with the dog's position at the two correct calls (green circles), and on one time axis the dog's distance to each target (green shading shows passes within 1 m), the coded handler gestures and commands, the dog's indications, and the handler's calls. (C) The chain from a hidden target to a call, with the number of the 218 targets that reached each stage (each target counted at the furthest stage it reached, so called targets count as passed and indicated).

**Table 1.** *Search environments, time limits, and hides, by environment and event.*

| | Outdoor area | | Indoor room | | Vehicles | |
|---|---|---|---|---|---|---|
| | **2025** | **2026** | **2025** | **2026** | **2025** | **2026** |
| Search area | 55.0 m² (approx. 7.2 × 8.4 m) | 50.6 m² (7.8 × 6.5 m) | 63.5 m² (approx. 8.4 × 9.2 m) | 59.3 m² (7.4 × 8.0 m) | 213 m² fenced area | 162 m² fenced area |
| Time limit | 3 min | 2 min | 2 min | 2 min | 4 min | 2 min |
| Targets | meth, cocaine | meth, cocaine | meth, cocaine | meth, cocaine | heroin (inside cargo van), cocaine (outside) | heroin (inside cargo van), cocaine (outside) |
| Distractors | hot dogs, dog biscuits, dog food, cat food, circulated and uncirculated currency | hot dogs, cheese, circulated and uncirculated currency | hot dogs, cheese, dog biscuits, circulated and uncirculated currency | hot dogs, cheese, dog biscuits, circulated and uncirculated currency | hot dogs, cheese, circulated currency, uncirculated currency (inside the van) | hot dogs (two hides), cheese, circulated and uncirculated currency (both inside the van) |

*Note.* Search areas were measured from the calibrated camera reconstructions and refer to the area teams could search. "Meth" = methamphetamine. Distractor lists are those present in the item layouts used for the analyses.

Each search contained two narcotics targets and several non-target distractors, which were chosen to resemble items that can elicit false alerts in real-world searches. We used distractor items of two different categories: highly attractive food items (hot dogs, cheese, dog biscuits, pet food), which is intrinsically attractive to dogs, and currency, which is not intrinsically attractive to dogs. However, because a vast majority of circulated banknotes are contaminated with trace amounts of target substances (Jourdan et al., 2013; Poupko et al., 2018), they can emit target odors (Furton et al., 2002). This can make circulated currency a potential confound in real-world searches because a dog's response to it may reflect a learned association between target odor and reward. Currency therefore represented objects that may have become associated with target odor, and uncirculated notes served as a control for the scent of currency itself, independent of the presence of narcotics. The circulated notes were ordinary circulated U.S. currency, to which no odor was added, and the uncirculated notes were new notes from a bank; each currency hide contained 30 to 80 bills. Targets and distractors were set by the professional trainer at least 30 min before the first search in each environment. Substances were handled with gloves and placed in containers (e.g., plastic bags, metal clips) in quantities ranging between approximately 10 and 100 g. Controlled substances were supplied by the trainer and by participating judges, who held all required licenses. Target and distractor locations were held constant within an environment and year; however, some hides in movable locations (e.g., cardboard boxes and tins) shifted slightly between teams despite being reset. We therefore localized every item separately for each trial (see Position tracking).

***Procedure***

All handlers were blind to the locations of targets and distractors. They were informed that each area contained exactly two targets and additional distractors. A search started when the handler crossed the designated start line and ended when the time limit expired or the handler declared the search complete. Handlers could choose whether to search on or off-leash and which leash to use. Most used a fixed-length leash (2025: 26 of 28 trials, 2 retractable; 2026: 67 of 81 trials, 10 retractable, 4 other). When a handler judged that their dog had found a target, the handler called an alert by raising a hand and/or verbally declaring the find. We refer to these handler alerts as *calls*, and to the dog's trained response at the target (e.g., a sit or a stare) as an *indication*. Judges recorded each call but gave no feedback on its correctness, so handlers did not learn the outcome of their calls during the search. At the end of a trial, handlers had the option to reward their dog at a known find placed past each area’s exit.

***Judging and scoring***

Every search was scored independently by expert judges (experienced detection dog trainers or law enforcement handlers). A panel of six judges scored all trials in 2025, whereas two judges scored each trial in 2026. Judges sat or stood at the edge of the search area. On a standardized sheet (Supplementary Figure S11), each judge recorded whether the team located each of the two targets, whether any false alert occurred during the search, and a total score from 0 to 100, from which points were deducted for false alerts, missed targets, and further performance categories. The false-alert entry was a single yes/no response for the whole search and did not record the number or location of false alerts. Judges watched the searches live, without video playback, and knew the hide locations. The searches were therefore single-blind, in that handlers did not know the hide locations, but the judges present did.

We averaged the judges' total scores, and classified a target as found or a trial as containing a false alert when at least half of the judges recorded the corresponding response (i.e., three judges in 2025 and either judge in 2026). Judge numbers differed across years with six judges per search in 2025 trials and two judges per search in 2026 trials. Judges' scores in 2025 were on average much higher (M = 65 vs 37). We therefore standardized the raw judge scores within each year without pooling them. To assess inter-judge agreement, we used Fleiss' kappa for six judges (Fleiss, 1971), Cohen's kappa for two judges (Cohen, 1960), and ICC for total scores (Shrout & Fleiss, 1979). Agreement was high for successful target location (kappa = .79 to .98) and the presence of a false alert (.83 and .87), and moderate-to-high on the total score (ICC(1) = .67 and .84; see Supplementary Table S3). We did not use the judges' sheets to classify calls due to the location not being noted down consistently (see below), but used them as an independent cross-check when available.

### *Video recording and behavioral coding*

All searches were recorded with fixed cameras at the edge of each search area, two to three per trial in 2026 and at least four in 2025, plus cameras inside the cargo van in the vehicle environment (models and placements in Supplementary Methods S1). Each environment was additionally captured with a handheld ZED 2i stereo camera (Stereolabs) as a metric three-dimensional scan, which served only as a geometric reference for calibration. Video recordings from each camera were synchronized to within one frame. We cut a separate clip per trial and camera that began 15 s before the handler crossed the start line; all times are expressed relative to that *(t* = 0, the start of the search). These clips formed the basis for our behavioral annotations.

Trained coders coded all trials in ELAN (Wittenburg et al., 2006) from the trial clips, with all camera angles of a trial available side by side, using a written ethogram (Supplementary Table S4). The coded variables were the trial, each handler call, the dog's indication at the called location (sit, stare, or other, coded as an interval), handler gestures, commands, praise, and periods off leash. Dog indications were thus defined with reference to the handler's calls and were coded with the calls in view, so nearly every call has a coded indication by construction; indications that the handler never called were coded only when a coder noticed them. Separate instances of a dog indication, a handler gesture, or praise less than 2 s apart were coded as one bout. Each trial was coded by one of five coders (four research assistants and the third author) in 2026 and by two coders in two passes in 2025 (trial boundaries and calls by the third author; then gestures, commands, praise, and off-leash periods), all blind to the locations of targets and distractors. To assess reliability, the third author re-coded 18 of the 2026 trials (six per primary coder) blind to the primary coding. Agreement on whether a behavior was occurring within a given second was good for handler gestures, praise, and dog indications, and moderate for commands (Cohen's kappa = .47 to .79; Supplementary Methods S1.8). Coded rates were higher in 2025 than in 2026 (gestures 7.9 vs 4.9, commands 4.8 vs 2.0 per minute), which could reflect coder calibration as well as differences between the two samples. All models therefore compare behavior within and not between events.

Because the coders saw the calls, indications that the handler did not call were checked separately. S.K. re-coded clips of all uncalled indications, of called indications, and of control moments without a coded indication, blind to which was which and with every person masked (Supplementary Methods S1.9). Following the bout rule of our ethogram, an uncalled indication was counted as part of a call when it ended no more than 3 s before a call at the same spot or when the dog stayed within 2 m of the spot until or after the call, and as a separate uncalled indication otherwise.

***Position tracking***

We reconstructed the positions of handlers and dogs with a custom Python pipeline that utilized open-source detection and pose models. Subsequently both were projected into metric ground coordinates for every frame of every trial (Supplementary Methods S1 gives each step with its parameters and validation). The lens (intrinsic) and pose (extrinsic) parameters of each camera were estimated against the ZED scan (Zhang, 2000; Triggs et al., 2000), and image points were mapped to the ground through homographies (Hartley & Zisserman, 2004) at the height of the handler's ankles and the dog's paws (held-out ground errors of 1 to 9 cm). Humans and dogs were detected with YOLO11x (Jocher & Qiu, 2024) and kept to one identity per subject. We operationalized the ground point as the midpoint between the handler's ankles (YOLO11x-pose) and the centroid of the dog's paws (DeepLabCut SuperAnimal-Quadruped; Mathis et al., 2018; Ye et al., 2024). We fused the positions from all cameras per frame and subsequently smoothed (Kalman, 1960; Rauch et al., 1965) and interpolated across gaps of up to 1 s (2 s at the vehicles). To control for some items moving slightly (e.g., cardboard boxes), every target and distractor was localized in every trial. Figure 1B shows these data for one search, and Figure 1C the chain from pass to indication to call. After fusion, the dog was tracked for a median of 89 to 100% of trial time outdoors and indoors and for 63% (2025) and 85% (2026) of trial time at the vehicles, where the dog was hidden behind vehicles for part of the search.

To locate and verify every handler call, we paired each call with the coded dog indication closest in time within 3 seconds. We then took the dog's position at the call as the median tracked position over the indication interval, and assigned the call to the nearest item within 1.5 m. This formed the basis for classifying a call as correct (on a target), or as a false call (on either a distractor or nothing nearby). Every handler call was subsequently verified by the authors on a clip showing two synchronized camera views with items labeled and dog and handler marked (Supplementary Methods S1.7). Where it differed from the coding by the live judges during the events (12 of 109 trials), the review was retained, because it rested on repeated viewing from several angles, including inside the van. Overall, the reviewed coding agreed with the judges' sheets on the number of targets found in 81 to 100% of the searches per environment and event, and the judges' mean score correlated with the coded number of targets found at $\rho = .81$ (Supplementary Results S2.6).

***Derived measures***

Using the fused tracks, item positions, and behavioral codes, we built a panel at 10 Hz for every trial. We operationalized a pass as a series of frames in which the dog's paw centroid was within 1 m of an item, and a pass as having produced an indication when a coded dog indication on that item started between 1 s before entry and 2 s after exit. The 1-m radius follows earlier work that counted passes closer than 1 m without an indication as missed opportunities (Jezierski et al., 2014) and scored an indication as correct when it occurred within 1 m of the target odor (Lazarowski & Dorman, 2014). From the panel we computed 13 search-process measures per trial, and we used several questionnaire responses on handler and dog (Supplementary Methods S1.10 and Supplementary Table S10).

***Statistical analysis***

Trials were nested in teams (three per team), and teams in years. Every model therefore included a random intercept for the team and the six environment-by-year cells as fixed effects, so that year absorbs everything that differed between the events (rooms and hide layouts, coders, judges, time limits, and the partial confound between year and professional background). Binary and count outcomes were fit as generalized linear mixed models (GLMM) and continuous outcomes as linear mixed models (LMM), except the false-call models, which were fit as generalized linear models (GLM) because their team variance was estimated at zero. We modeled four outcome families: (1) whether a target was found, per target (binomial; 218 target-trials), (2) the number of false calls per trial (Poisson, with the log of trial duration as offset, turning effects into rate ratios per minute of searching), (3) the time to the first correct call (Cox proportional hazards; Cox, 1972), stratified by cell and censored at the end of the trial, with hazard ratios above 1 meaning sooner finds, and (4) whether a pass produced an indication (binomial; 1,728 passes before the item was called). Because the process measures were strongly right-skewed, we transformed continuous predictors to rank-based normal scores (inverse-normal transformation of ranks). We report effects per SD of normal scores, as odds, rate, or hazard ratios (OR, RR, HR) with 95% confidence intervals.

Predictors were screened one at a time within two predefined sets, team characteristics (seven survey predictors and professional background) and search process (13 measures), with Benjamini-Hochberg false-discovery-rate q-values within each screen (Benjamini & Hochberg, 1995). We treat $q < .10$ as the threshold for interpretation and $p < .05$ with a larger q as suggestive. The predictors that met this threshold were then entered together in one model. To test for location-specific cueing, handler behavior was compared between targets and distractors on passes without an indication, and between indications that became correct and false calls, with the same mixed-model structure. To test whether calling changed as the time limit approached, we compared the last 30 s of a search with the time before, in the 58 searches that ran to within 5 s of the limit.

We partitioned the variance of each trial-level outcome between event, environment, professional background, team, and trial, and quantified team consistency as the intraclass correlation of each measure across a team's trials. Model details, robustness checks, and the replication of the key models within each event are given in Supplementary Methods S1.11 and S1.12.

### ***Software***

For tracking, we used Python 3.12 with PyTorch (Paszke et al., 2019), Ultralytics YOLO11 (Jocher & Qiu, 2024), DeepLabCut 3.0 (Mathis et al., 2018; Nath et al., 2019), OpenCV (Bradski, 2000), NumPy (Harris et al., 2020), and SciPy (Virtanen et al., 2020) on a workstation with an NVIDIA GeForce RTX 5090 GPU. The environmental scans were processed with the ZED SDK 5.2 (Stereolabs). All statistical analyses were run in R 4.3.3 (R Core Team, 2024) with lme4 1.1-35.1 (Bates et al., 2015), lmerTest 3.1-3 (Kuznetsova et al., 2017), survival 3.5-8 (Therneau, 2024), sandwich 3.1-0 (Zeileis et al., 2020), and lmtest 0.9-40 (Zeileis & Hothorn, 2002). The analysis tables were built in Python 3.12 with pandas (McKinney, 2010), and the figures were drawn in R with ggplot2 3.4.4 (Wickham, 2016), patchwork 1.2.0 (Pedersen, 2024), and ggrepel 0.9.5 (Slowikowski, 2024).

## Results

### ***Search outcomes and the range between teams***

Table 2 summarizes the 109 searches. Handlers made 157 calls, of which 92 (59%) were on a target and 65 (41%) were false alerts (53 on a distractor and 12 with no distractor nearby). Teams found 92 of the 218 targets (42%; per year: 2025: 55%, 2026: 38%). Across the two years, both targets were found in 19% of trials, one in 46%, and neither in 35%, with at least one false call occurring in 50% of trials (2025: 39%, 2026: 54%). Figure 2 shows the composition calls per trial and the share of targets found in each environment and in each event.

**Table 2.** *Search outcomes by environment and event.*

| | **Outdoor** | | **Indoor** | | **Vehicles** | |
|---|---|---|---|---|---|---|
| | **2025** | **2026** | **2025** | **2026** | **2025** | **2026** |
| Searches | 10 | 27 | 9 | 27 | 9 | 27 |
| Handler calls per search | 1.20 | 1.30 | 2.00 | 1.44 | 1.67 | 1.41 |
| Correct calls per search | 0.80 | 0.56 | 1.44 | 0.74 | 1.11 | 0.96 |
| False calls per search | 0.40 | 0.74 | 0.56 | 0.70 | 0.56 | 0.44 |
| Share of calls correct | .67 | .43 | .72 | .51 | .67 | .68 |
| Targets found, % | 40 | 28 | 72 | 37 | 56 | 48 |
| Targets found within 120 s, % | 30 | 26 | 72 | 37 | 28 | 44 |
| Searches with at least one false call, % | 40 | 63 | 33 | 63 | 44 | 37 |

*Note.* Time limits were 3, 2, and 4 min (outdoor, indoor, vehicles) in 2025 and 2 min in all three 2026 environments. Six judges scored each 2025 search and two each 2026 search; judge scores are reported in Supplementary Table S3.

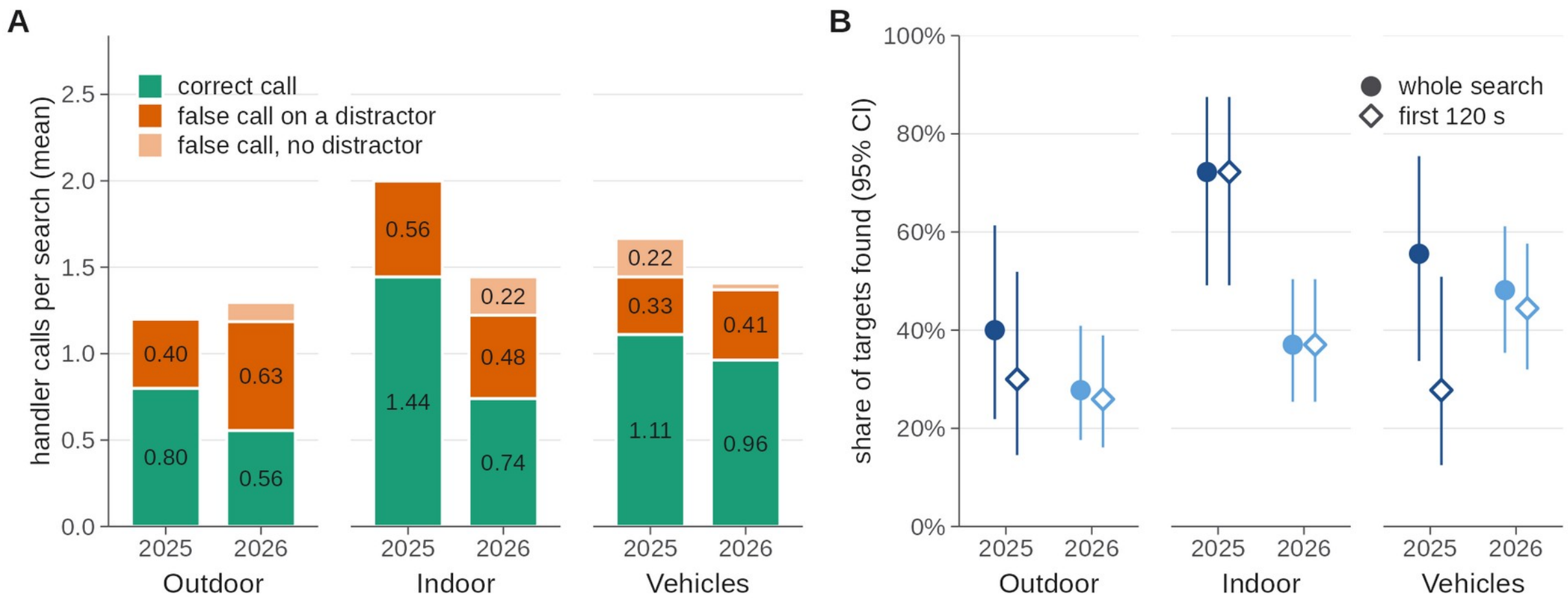

**Figure 2.** (A) Handler calls per search by outcome class, and (B) the share of targets found over the whole search and within the first 120 s (95% Wilson confidence intervals), by environment and event. Calls are hand-reviewed on video.

The relative success of target location across environments differed between years, likely because 2025 teams had one to two more minutes to search outdoors and at the vehicles (outdoor from 30% of targets within 120 s to 40% over the whole search, vehicles from 28% to 56%; Figure 2B; Supplementary Figure S1). Within the first 120 s, the outdoor area and the vehicles yielded the fewest targets in 2025 (30% and 28%) and the outdoor area the fewest in 2026 (26%), whereas the 2025 room stood out (72%). With the team random intercept taken into account, however, the environment differences in finding within each year were modest (likelihood-ratio tests: 2025 $p = .066$, 2026 $p = .063$), the rate of false calls per minute did not differ between environments within either year ($p = .40$ and $.31$), and 2026 teams tended to find fewer targets than 2025 teams over the whole search (OR = 0.40, 95% CI [0.15, 1.10], $p = .077$) but not within the 120-s window (OR = 0.67, [0.26, 1.74], $p = .42$). The hide itself seemed to matter more than the environment or the substance. The same substance in the same environment was found in 70% of the searches in one year and 19% in the next, and the substance had no significant effect of its own (Supplementary Results S2.1).

Figure 3 shows every team's outcome in every environment. The 36 teams that completed all three searches found between zero and five of their six targets. Five teams missed all six targets, six missed only one, and no team found all six. Sixteen of the 36 teams found at least one target in every environment, six made no false call in their three searches, and only two found four or more targets without making a false call. The teams with the most finds were not free of false calls (nine of the 11 teams with four or more finds made at least one), but they made fewer of them (13 false calls among these 11 teams as compared to 52 among the remaining 25; Spearman $\rho$ between targets found and false calls = −.45).

The best searches were fast. One 2026 team called its first outdoor target 7 s after crossing the start line, and at the vehicles it called its first target after 9 s and the second after 36 s. Both targets of a search were also found within 41 s in the 2025 room and within 66 s outdoors in 2025. Over all 71 searches with a find, the first target was called after a median of 65 s (IQR 41 to 95 s), and in the 21 searches in which both targets were found, the second was called after a median of 91 s.

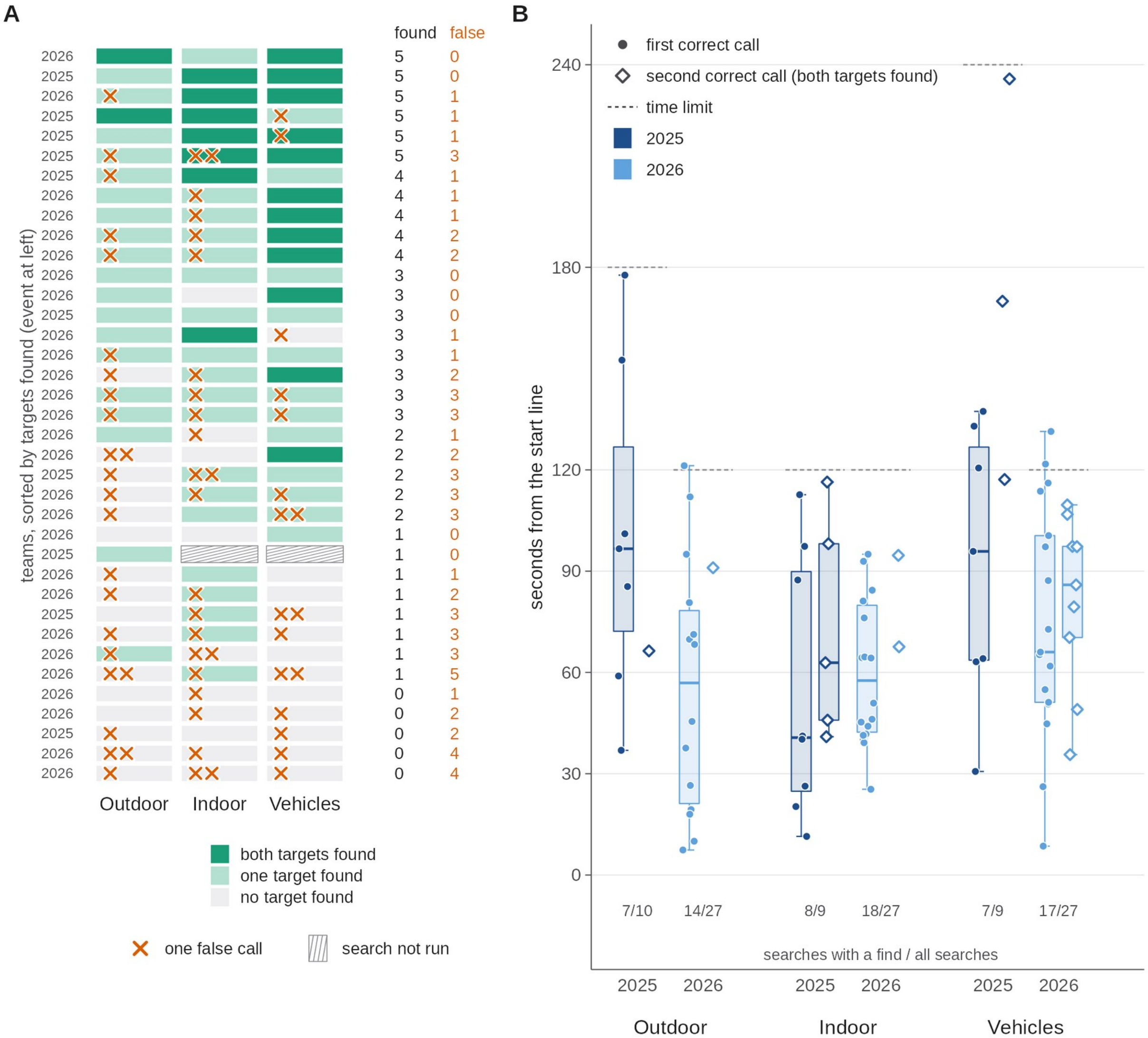

**Figure 3.** Team profiles. (A) Targets found (green fill) and false calls (crosses) per team and environment, one row per team, sorted by the number of targets found over the three searches; the event of each team is given at the left. (B) Time from the start line to the first correct call (dots) and, where both targets were found, to the second (diamonds), per environment and event, with box plots for cells with at least four searches; dashed lines mark the time limits, and the numbers give the searches with a find out of all searches in that cell.

### *Where finds are lost*

Dogs came within 1 m of 92% of the targets (201 of 218) and covered the search areas densely, with the highest occupancy along objects at the edges of the areas and around the vehicles (Supplementary Figure S2). Of the 17 targets that the dog never approached within 1 m, half (8 of 17) were the cocaine hide outside the 2026 vans, which lay where the dog was occasionally out of view of the exterior cameras. Therefore, 8% unreached is an upper bound. Of the 201 approached targets, 91 were called and 110 were not. The 111 approached targets without a coded dog indication (108 never called and 3 called without one) had been passed a median of 3 times (IQR 2 to 4), and the dog had spent a median of 7.7 s within 1 m of them over the search (IQR 3.8 to 11.6 s). Targets that were eventually indicated received a median of 5.9 s within 1 m before the first indication (IQR 3.5 to 9.9 s). Therefore, missed targets had not been passed more briefly than found ones.

***Indications per pass***

The 1,728 passes made before an item was called (542 at targets and 1,186 at distractors) can be treated as detection trials with known answers (Figure 4). A pass at a target produced a dog indication 17.3% of the time and a pass at a distractor 3.7% of the time. The odds of an indication were 5.8 times higher at a target (95% CI [4.0, 8.6], $p < .001$). Dogs therefore discriminated between targets and distractors in every environment and year. But they responded on only about one in six target passes. Prior to a target being found, it had been passed a median of 2 times without an indication (IQR 1 to 3). Only 4% of finds occurred on the first pass. By item, indication rates were 21% of passes at methamphetamine, 17% at cocaine, 13% at heroin, 7% at hot dogs, and 5% or less at the other distractors (Figure 4B). The discrimination did not depend on the pass radius (odds ratio 4.8 at 0.5 m and 5.3 at 1.5 m; Supplementary Table S7) and held in each event alone (13.9 in 2025 and 4.3 in 2026; Supplementary Table S8). Notably, passes without an indication were brief. At targets, the dog remained within 1 m for a median of 1.7 s (IQR 0.9 to 3.6), and 28% of target passes and 26% of distractor passes lasted less than 1 s. Per team, the median hit rate per target pass was 17% (IQR 6 to 31%) and the median false-alarm rate per distractor pass 3% (IQR 0 to 6%). Rates can be estimated even for teams that never found a target. Seven of the 37 teams never indicated on a target pass, whereas the six teams with five finds indicated on 24 to 56% of theirs.

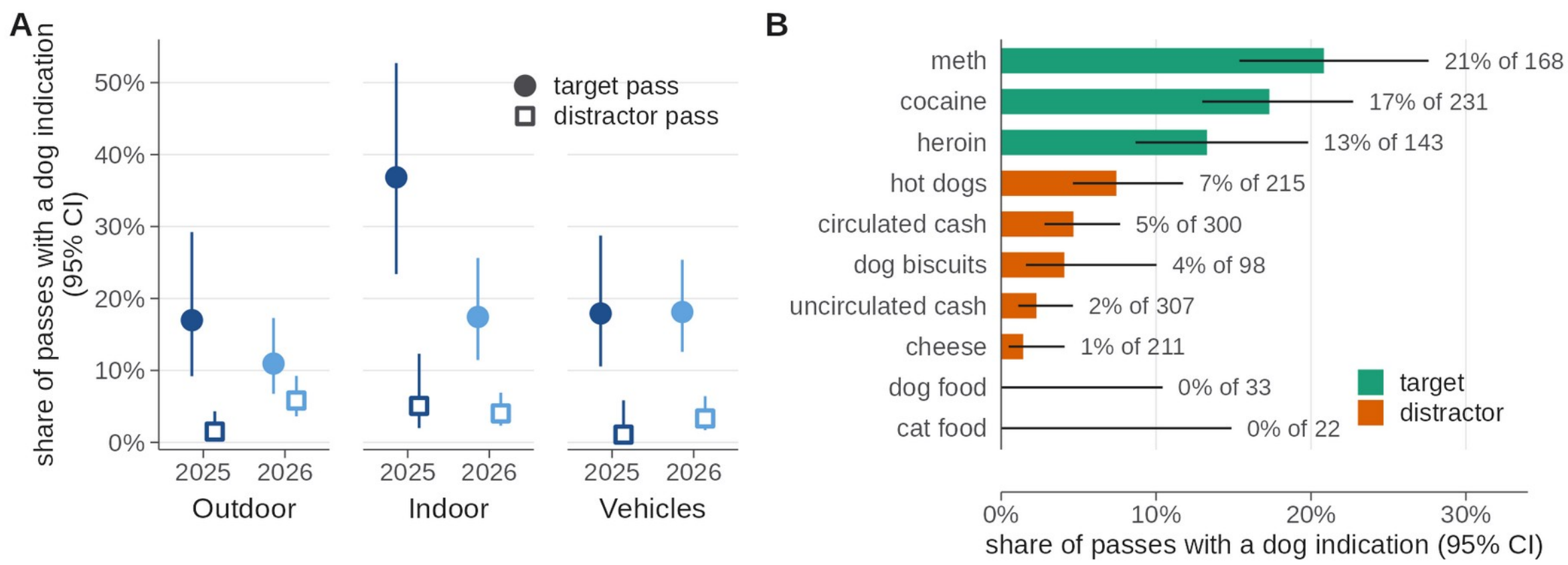

**Figure 4.** Passes as detection trials. (A) Share of passes within 1 m that produced a dog indication, at targets and at distractors, by environment and event, with 95% confidence intervals. (B) The same share by item, both events pooled, with the number of passes (green: targets; orange: distractors). We excluded passes that happened after the item had been called.

### *The handler's role in the outcome*

If handlers had cued their dogs at specific hide locations, they should have behaved differently near targets than near distractors before the dog responded. On the 1,590 passes without an indication, they did not (Figure 5A). Around the dog's entry into a pass (3 s before to 1 s after), handlers pointed during 3.5% of the time at targets and 3.6% at distractors, gestured in any way during 9.5% and 10.6%, and gave a verbal command during 2.5% and 2.2%. They moved at 0.61 and 0.66 m/s on entry and stood 2.07 m from the item in both cases, and the passes lasted 2.8 and 3.0 s on average (all $p \geq .19$).

The moments before the dog indicated might provide a more informative comparison. In the 10 s before an indication, handlers gestured during 14% of the time before indications that became correct calls and 7% before indications that became false calls ($p = .023$; Figure 5B), a difference with the same direction in both events but significant only in 2026 ($p = .036$; 2025 $p =$

.43; Supplementary Table S8). The same direction was true for pointing (3.2% vs 0.8%, $p = .13$), whereas commands did not differ (about 1% throughout). In both cases, the dog slowed from about 0.7 to 0.2 m/s over the 5 s before the indication (Supplementary Figure S3). Handlers' gesturing before correct calls declined toward the indication, consistent with a handler presenting or detailing a location and then allowing the dog to assess. We therefore found no evidence that handler movement, gestures, or commands preceded false calls more than correct calls.

Once a dog indicated, its handler almost always called the item. Of the 91 targets with a coded dog indication, 89 were called, and every coded indication at a distractor but one, at the cheese, was eventually called as well. The blind re-coding confirmed that the uncalled indications were real dog behavior (an indication was seen in 90% of the judgeable uncalled clips, 86% of the called clips, and 14% of the control clips), but most belonged to the same bout as a call. Of the 69 uncalled indications in the analyzed teams, 55 merged into a called bout and 4 were dropped because the blind coder saw no indication, which left 8 bouts (10 coded indications in 8 searches) that were separate from any call. In five of these, the handler called the same item on a later visit, 10 to 57 s afterward (four targets and once the hot dogs), and in one the dog indicated again at a target that had already been called. Two were never called, at the cocaine outside the 2025 vans and at the cheese. Because indications that were never called were coded only when a coder noticed them, it is possible that they undercount the indications that handlers missed.

Calling became more frequent as the time limit approached. In the 58 searches that ran to the time limit, handlers made 1.03 calls per minute in the last 30 s against 0.34 before. Correct calls ($RR = 2.33$, [1.39, 3.92], $p = .001$) and false calls ($RR = 3.73$, [1.76, 7.89], $p = .001$) both increased, so the share of calls that were false rose only from 44% to 57% (Fisher's exact test p =

.46). The dogs did not pass items more often late in a search, but distractor passes more often produced an indication (8.3% vs 1.5%; OR = 5.90, [2.37, 14.73], p < .001), whereas the rate at targets changed less (11.1% vs 9.4%, p = .46; Figure 5C). The increase at distractors came mainly from the 2026 event (Supplementary Table S9). The 12 false calls with no distractor nearby also came late (median 95 s, against 60 s for false calls on a distractor and 69 s for correct calls).

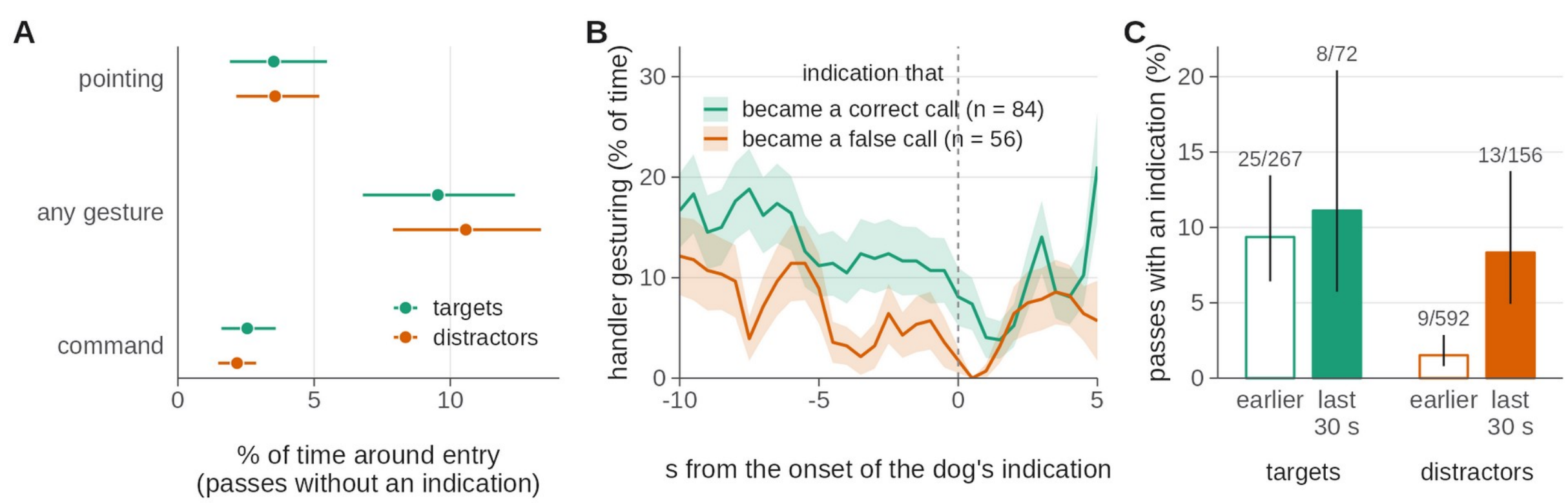

**Figure 5.** The handler's part. (A) Handler behavior around the dog's entry into a pass (3 s before to 1 s after) on the 1,590 passes without an indication, at targets and at distractors, as the share of time with pointing, any gesture, or a verbal command (mean with 95% bootstrap interval over teams). (B) Handler gesturing from 10 s before to 5 s after the onset of a dog indication, for indications that became correct and false calls (mean ± SE, 0.5-s bins). (C) Share of passes that produced a dog indication, earlier in the search and in its last 30 s, at targets and at distractors, in the 58 searches that ran to the time limit (95% Wilson intervals; numbers give indicated passes out of all passes).

### ***The team and the individual search***

Figure 6A shows the variance of the trial-level outcomes attributable to the event, the environment, professional background, and the team. For the number of targets found, the environment (within event) accounted for 7% of the variance, the event year for 5%, and handler's professional background for 5%, whereas the identity of the team accounted for 33% (residual ICC = .39, permutation p < .001), with the remaining 50% varying from trial to trial within teams. The environment accounted for 25% of the variance in the mean judge score and the team for 22% (ICC = .31, p = .008). For the number of false calls, neither the environment (3%) nor the team (17%, ICC = .17, p = .06) accounted for much variance, 79% varied from trial to trial. Of the 37 teams, seven made no false call over their searches, 18 made one or two, and 12 made three or more (permutation test of the variance across teams, p = .61; Supplementary Figure S4).

Accordingly, a team's three searches were moderately consistent in the number of targets found but not in the number of false calls (Figure 6B). After removing environment-by-event means, intraclass correlations across a team's searches were .43 for targets found (bootstrap 95% CI [.21, .61]) and .37 for the standardized judge scores ([.14, .56]), as compared to .18 for false calls ([−.03, .38]) and .18 for whether any false call occurred ([−.02, .39]), and the pattern held in each event alone (Supplementary Table S8). Teams also had recognizable search styles. The share of time the handler gestured (ICC = .43, [.22, .64]) and the dog-handler distance (.40, [.33, .66]; median 1.12 m) were moderately consistent across environments, followed by the dog-ahead share (.34), dog speed (.33), and handler speed (.25), but none of these style measures predicted finding (Figure 7).

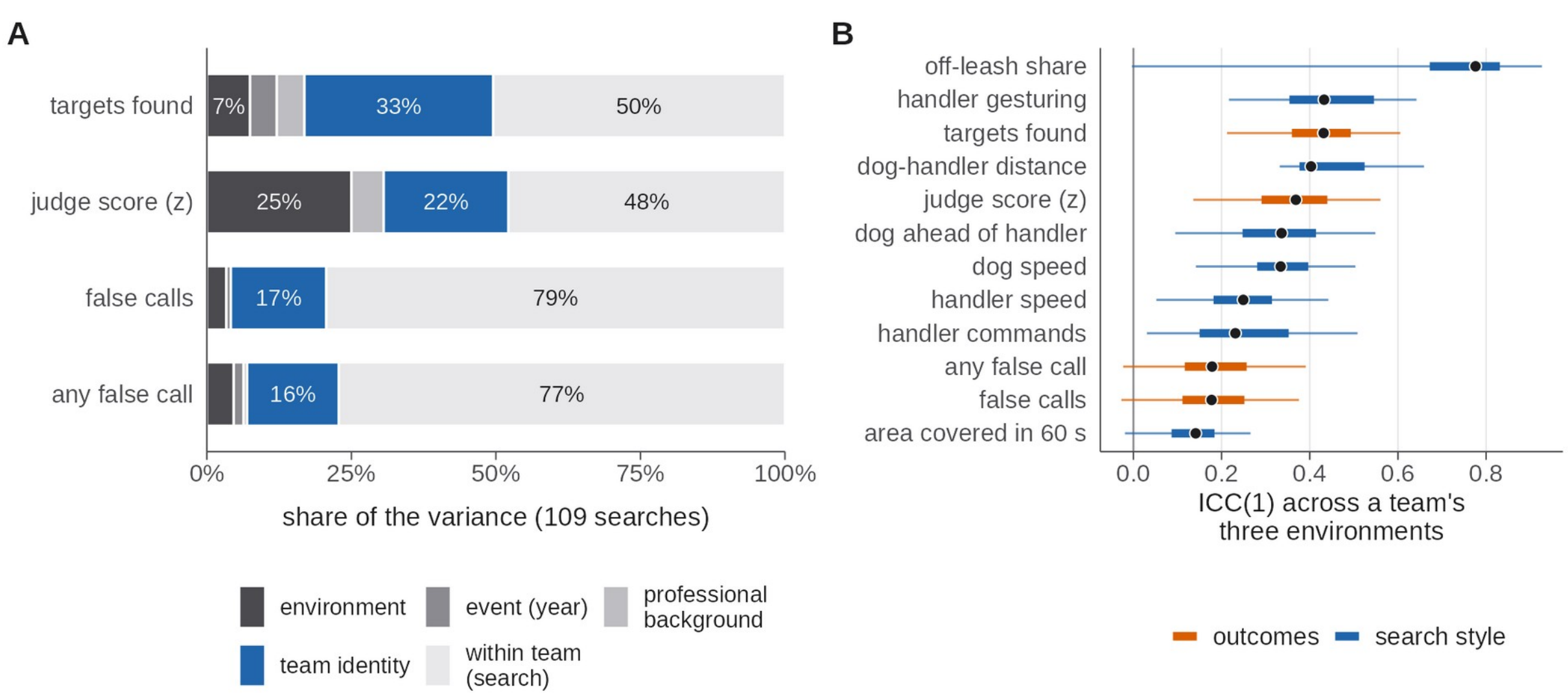

**Figure 6.** The team and the individual search. (A) Share of the variance in each trial-level outcome across the 109 searches attributable to the event, the environment within the event, professional background, team identity (ICC of the residuals), and variation between a team's searches. (B) Team consistency, both events pooled: ICC(1) of each measure across a team's three environments after removing environment × event means (dot), with the interquartile range (thick bar) and the 95% interval (thin line) of a bootstrap over teams (2,000 resamples). Orange: outcomes; blue: search-style measures.

False calls instead concentrated on particular items. Across all trials, hot dogs elicited 19 false calls, circulated currency 17, uncirculated currency 9, cheese 4, dog biscuits 3, and dog food 1. Additional 12 calls were made with no distractor nearby (Supplementary Table S6 and Figure S9). Hot dogs were the dominant outdoor lure in 2026 (12 of the 20 outdoor false calls), and circulated currency the dominant indoor lure (8 of 19). Per pass, food and currency elicited false calls at the same rate (4.0 vs 3.8 per 100 passes; OR = 1.01, [0.55, 1.87]), whereas within the food class, hot dogs elicited four times more calls per pass than the other food items (OR = 4.26, [1.57, 11.60], $p = .005$). Supplementary Results S2.2 gives the rates per item and the calls at no distractor and inside the van.

***What predicts finding and false calls***

Figure 7 shows the predictor screens for finding a target and for false calls. Of our 13 process measures, only the handler's median speed during the search phase predicted whether a target was found, with faster handlers finding more targets (OR = 1.91 per SD of normal scores, 95% CI [1.19, 3.07], $p = .007$, $q = .092$). The dog's speed was more weakly associated with finding (OR = 1.35, [0.93, 1.98], $p = .12$, $q = .51$), and this association disappeared once handler speed was entered in the same model (handler OR = 2.32, [1.19, 4.52], $p = .013$; dog OR = 0.79, [0.44, 1.39], $p = .41$), so the effect is likely carried by the pace of the team. The two speeds were correlated ($\rho = .62$), as expected given that 82% of searches were conducted entirely on leash, and these models cannot establish which partner sets the pace. Because every model compares searches within the same environment and event, the size of the search areas does not explain the effect, which had a similar size in each event (2025 OR = 2.10, $p = .13$; 2026 OR = 1.85, $p =$

.028). To address reverse causation (teams speeding up after a find), we measured only before the first dog indication of the trial or only in the first 30 s. The association kept its direction but was weaker and no longer significant on its own (handler OR = 1.43, [0.98, 2.08], $p = .065$, and OR = 1.44, [0.96, 2.16], $p = .076$). It was primarily a within-team effect, as teams found more targets in searches in which they moved faster than their own mean across environments (within-team OR = 2.38, [1.31, 4.32], $p = .004$), whereas teams that were generally faster were not reliably more successful (between-team OR = 1.33, [0.65, 2.73], $p = .43$). No other process measure predicted finding (all $q \geq .51$).

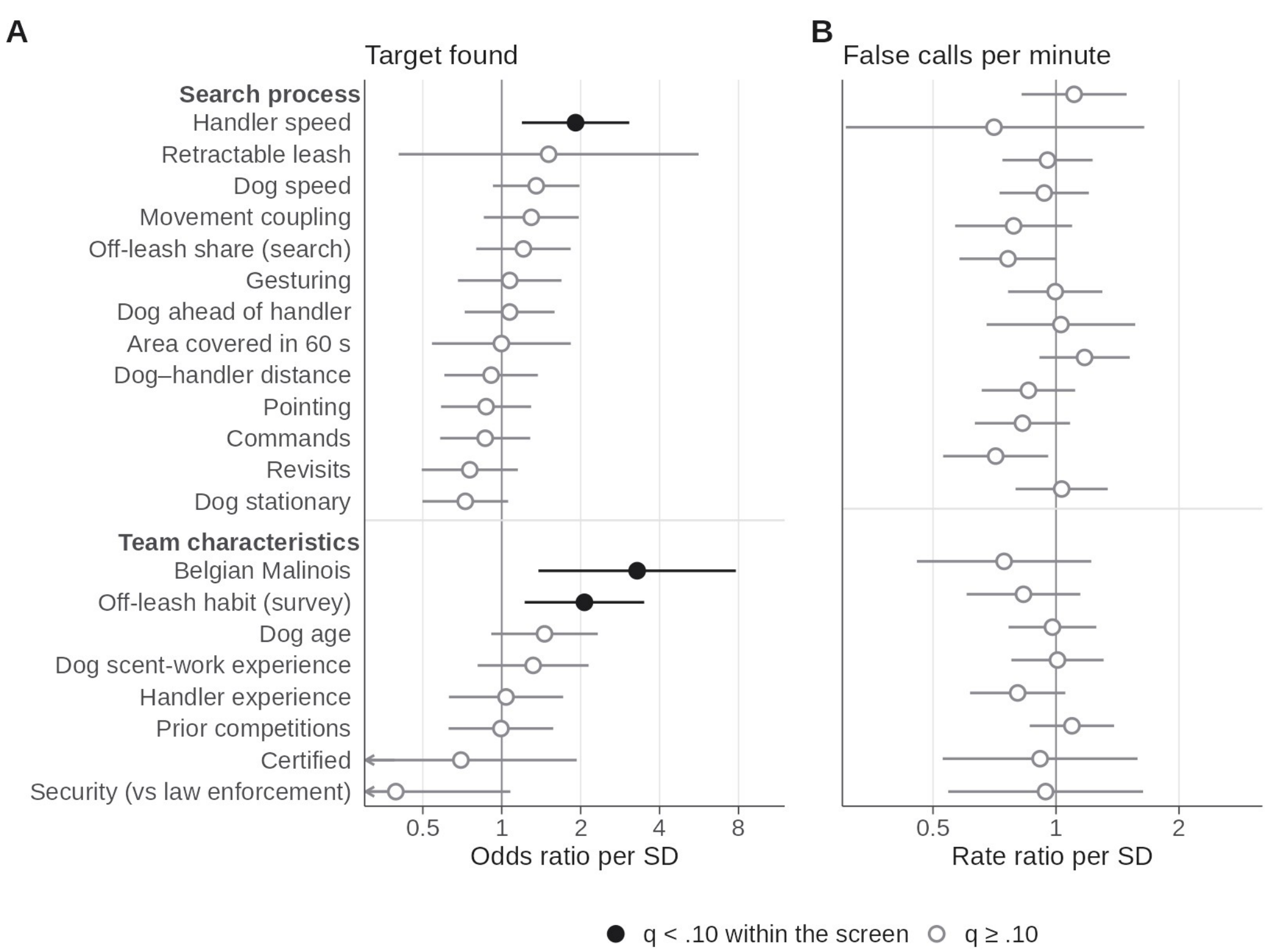

**Figure** 7. Predictor screens for finding a target (A; GLMM with a team random intercept; odds ratio) and for false calls per minute (B; GLM with search duration as exposure; rate ratio), one predictor per row. All models are adjusted for environment × event; effects are per SD of normal scores for continuous predictors, with 95% CI. Filled markers: $q < .10$ within the screen (professional background: $p < .05$). The screen for the time to the first find is in Supplementary Figure S10.

Among team characteristics, characteristics of the dog were the main predictors of finding. Belgian Malinois found more targets than other breeds (OR = 3.28, [1.38, 7.81], $p = .007$, $q = .025$), and teams whose handlers reported that their dog searched off leash more often found more targets (OR = 2.07 per step of the three-point scale, [1.22, 3.49], $q = .025$). Off-leash search duration during the trial had no effect. The survey item might therefore index a general style of the team rather than behavior on the day. When breed, off-leash habit, and handler speed were entered together, all three remained associated with finding (ORs 2.1 to 4.8, all $p \leq .001$). The same two characteristics, together with dog age and scent-work experience, were associated with finding the first target sooner (Supplementary Results S2.7). Handler experience, certification, and prior competitions did not predict finding (all $q \geq .68$), and handler experience was only weakly associated with fewer false calls per minute (RR = 0.81 per SD of log months, [0.62, 1.05], $p = .11$). Security teams tended to find fewer targets than law enforcement teams (OR = 0.41, [0.16, 1.00], $p = .051$), but because they searched after the law enforcement teams in every 2026 environment, this contrast is confounded with search order (Supplementary Results S2.5).

No process measure predicted false calls at the paper's threshold. Revisits were associated with fewer false calls per minute (RR = 0.71 per SD, [0.53, 0.96], $p = .024$), and so, more weakly, was handler gesturing (RR = 0.76, [0.58, 1.00], $p = .053$), but neither survived correction within the screen ($q = .32$ and $.34$). Dog-handler distance, the two speeds, pointing, commands, the off-leash share, and leash type were unrelated to false calls ($p \geq .16$).

## Discussion

We analyzed 109 searches by 37 professional narcotics detection teams in three realistic environments as synchronized records of handler and dog positions, their communication, and dog behavior at each target and distractor. Teams found 42% of the targets but made a false call in half the searches, with wide differences between teams. Finds were missed mainly when the dog was close to a target but did not indicate, rather than due to incomplete coverage or because handlers failed to call an indication. We found no difference between handler behavior near targets or distractors before the dog responded. Calls of both kinds became more frequent as the trial time limit approached. Finding was a moderately consistent property of a team and was associated with the pace of the search, whereas false calls varied from search to search and clustered on specific distractors.

***Performance under realistic conditions***

Our teams found fewer targets and produced more false alerts than in the studies of certified teams of Rice and Velasco (2021, 2023) or the police dogs of Jezierski et al. (2014). However, these prior studies measured finds per search rather than finds per target and also tested under conditions that favor the teams (familiar sites and procedures, no distractors placed to attract the dog, single-blind testing). Our searches combined several features that have each been associated with lower performance on their own, namely unfamiliar sites and search-area sizes (Rutter et al., 2021), vehicles (Jezierski et al., 2014), and field conditions (Karpinsky et al., 2025), together with short time limits, clutter, and attractive distractors. The difference in performance is unlikely to be due to differences in scoring because the judges watching live searches agreed with the video review in more than four of five trials. Our results are also unlikely due to a sample of weak teams, because the handlers were certified and working professionals.

The same conditions also revealed the effectiveness of the most successful pairs, which found five of their six targets and called their first target within seconds of the start line (Figure 3). These are not the marks of teams that cannot detect. Rather, they show what affordances are available to a team that is actively engaged with an odor source. In addition to ability, lower performance in these conditions could plausibly be attributed to novelty of the environments, their layout, trial procedures such as the fixed time limit, the presence of judges and other teams. Additionally, hides whose odor availability may have differed between years due to environmental conditions such as weather (the same substance in the same environment was found by 70% of teams in one year and 19% in the next). These results should therefore not be read as estimates of real-world accuracy, which depends on base rates and on the many searches in which no target is present. They rather suggest that certification pass rates may not be reliable estimates of performance in real-world searches. Once the time limits were equalized, the open outdoor area was the most difficult environment in 2026 and, together with the vehicles, in 2025. A room or the inside of a van partly contains an odor plume, whereas hides on an open lot are exposed to wind, which strongly shapes a dog's search movements (Jinn et al., 2020; Mejia et al., 2024).

***Where finds were lost***

The outcomes of the individual targets argue against the most obvious explanation for the low find rate, namely that coverage was limiting. In searches lasting about 2 min over 50 to 210 m², the dog came within 1 m of 92% of the targets, and those on which it never indicated had been passed about three times, for a cumulative 8 s within 1 m, no less than the targets that were eventually indicated had received before their first indication. More opportunities may still have mattered. With about one indication in every six target passes, each further pass allows for another chance of a find, as the additional finds in the extra minutes of 2025 suggest. Nor was the handler's recognition of the response limiting, because handlers almost always called an indicated target. This leaves the pass on which the dog came within 1 m of a target but did not respond as the step at which most finds were lost. Treating passes as individual detection trials gives the free search what laboratory line-ups have, a hit rate and a false-alarm rate for every team on the same scale, regardless of whether the team ever found a target. The dogs discriminated between targets and distractors, with the best teams indicating on up to half of their target passes, and although the per-pass rates rest on codes by research assistants, the discrimination held at pass radii from 0.5 to 1.5 m and in each event, which suggests that the qualitative pattern is robust.

The pass cannot tell a dog that did not perceive the odor from one that perceived it and did not respond. In an exploratory check, dogs slowed below 0.2 m/s for at least 0.5 s on 24% of the passes without an indication, at targets and at distractors alike, and these pauses lasted about as long at targets as at distractors (median 2.1 vs 2.3 s). Because true and false negative responses differ in sniffing behavior (Concha et al., 2014) and odor responses begin before the trained response (Martvel et al., 2025), a future analysis could record head height, sniffing, and orientation on unproductive passes and possibly measure airflow at the hides.

***The handler in the loop***

A score sheet cannot tell whether a handler called an alert the dog never gave or if the dog alerted in response to the handler's expectations (Lit et al., 2011). Our design records the dog's indication and the handler's call as separate events and locates both, so that, in principle, this distinction can be made for every call. Because handlers were blind to the hide locations, the absence of target-specific handler behavior on passes without an indication is what the design predicts without evaluator cues, and it speaks against handlers having inferred hide locations from other sources, such as the judges. Handlers also gestured less, not more, before false calls than before correct calls, so the false calls in this setting were not preceded by the kind of handler involvement that a *Clever Hans* account would predict.

Handlers called nearly every indication their dogs gave, at targets and at distractors, although in five cases only after the dog indicated there again on a later visit. Some handlers thus seem to have waited for a second indication before committing to a call. With two targets known to be present, calls of both kinds became more frequent in the last 30 s of a search, and the 12 calls with no distractor nearby also came late. This late rise might reflect handlers who still expect a target and accept a weaker response, dogs that indicate more readily, or both, and our data cannot separate these sources. Whether handler belief cues the dog when the handler holds a belief about an exact location, as at the marked locations of Lit et al. (2011), remains unanswered by our design.

***Search pace, dog traits, and handler experience***

The only feature of the search process that predicted finding within our data was pace. Because the association was mostly within teams (the same team did better in the environments in which it moved faster, whereas habitually faster teams were not reliably better), it might reflect a state of the search rather than a trait of the team. It kept its direction, although weaker and not significant on its own, when speed was measured only before the first indication or only in the first 30 s, so this might not be a mere consequence of finding. A search that stalls might be one in which the dog is not on odor and the handler has not yet decided where to go next. Alternatively, handler pace might track difficulty, arousal, or fatigue. Therefore, this association is a process marker and not advice for handlers to walk faster. Search styles, such as the dog's working distance and the amount of gesturing, were consistent properties of teams but did not account for differences in success.

Characteristics of the team that were associated with finding targets were primarily related to the dog. Belgian Malinois found more targets, and teams whose handlers reported that their dog habitually searched off leash found more targets. Breed, off-leash habit, and handler speed each remained associated with finding when entered together. Older and more experienced dogs found their first target sooner. These are team-level predictors estimated from 37 teams and should be replicated in larger samples. Previous findings on breed are mixed (Hall et al., 2015; Jezierski et al., 2014; Rice & Velasco, 2023). In a professional sample breed likely indexes selection, procurement, and training as much as olfaction. A habit of off-leash searching might index training in which the dog is given more independence in where to search, or dogs that their handlers trust to work at a distance. Given that off-leash searching within our study did not predict finding targets, the effect is unlikely to be due to the leash itself. The handler's experience, certification, and competition history did not predict finding, and the tendency of more experienced handlers to make fewer false calls was not significant.

### *False calls depend on the individual search and on specific items*

The number of targets found was a moderately consistent property of a team, whereas false calls were not (ICC = .18, with an interval that includes zero). The distribution of false calls across teams was indistinguishable from a random process with the observed rates. Whether a team produced a false call depended mainly on the individual search, on the distractors it contained, above all hot dogs and circulated currency, and, as discussed above, on how much time was left. Food and currency drew false calls at the same rate per pass, but possibly for different reasons. A call on food might reflect the dog's interest in the item rather than an odor decision (hot dogs drew four times more calls per pass than the other food items), whereas a call on circulated currency might be an odor decision, given the trace contamination of banknotes, or a learned association between currency and reward (see Method). A single yes/no false-alert entry per search cannot capture these item-specific differences, whereas per-pass rates can. False alerts may therefore be more usefully studied and trained as responses to specific items in specific contexts (Caldicott et al., 2024; Gazit et al., 2005; Porritt et al., 2015) than as a trait of a team, and food and currency should be part of blind training.

***The team as a cognitive system***

If the dog's response at the hide is the stage at which most finds were missed, one might conclude that detection in this setting was a dog rather than a handler problem. However, our data do not support that reading. Of the criteria that Amon and Favela (2019) proposed for treating a human-dog dyad as one cognitive system, our data speaks to agency on both sides. The dog was ahead of the moving handler for less than half of the search phase in most environments (a median of 28 to 53%, depending on the environment and event). In a professional search, a handler is not simply following its dog. Instead, the handler determines where the dog is, for how long, and how many times it returns to a location, and decides whether a response becomes a call. The dog in turn has access to the odor and decides whether to indicate. Our data support that successful searches are the result of these two sets of decisions (e.g., the pace effect, the three passes at a typical missed target, the rise in responding late in a search). The same location affords the dog an odor to sample and the handler a place to present. A search succeeds when the two affordances are taken up at the same place and time. Therefore, locating the bottleneck at the dog's response points to where in the interaction the loss becomes visible and not which partner is solely responsible for it.

### *Limitations*

Our study has several limitations. First, dog indications were coded with the handler's calls in view, so nearly every call has a coded indication by construction. Therefore, some subtle indications that handlers never called may have gone uncoded. Coders agreed on dog indications with kappa = .61. Some of our computer vision tracking coverage was incomplete at the vehicles, especially in 2025. We did not record wind or airflow, and our search areas were small enough to be covered several times within the time limit. In many real-world search scenarios in larger areas, coverage may well be the limiting step. Handlers knew that each area held two targets, which may have contributed to the late rise in calling. In the current study, we measured handler movement, gestures, and commands but not gaze, body orientation, or the subtle leash signals that a dog might perceive. Subtle signals that we missed might carry meaningful information, such as the tail movements documented by Martvel et al. (2025) that even experienced observers miss. The judges also knew the hide locations, and although the similarity of handler behavior at targets and distractors makes a strong influence of judge cues on the handlers unlikely, some dogs might nonetheless have taken cues from the judges.

### *Conclusion*

Professional narcotics teams searching realistic environments with clutter and attractive distractors covered the search areas thoroughly yet found fewer than half of the targets, and performance ranged widely, from teams that found none of their six targets to teams that found five and called their first target within seconds. Following each target through the search showed where finds were missed. Dogs came close to nearly every target, and handlers called nearly every indication their dogs gave, but on most passes at a target the dog did not respond. Handlers did not behave differently near targets than near distractors, yet the team as a whole responded more readily as the time limit approached. Finding was a moderately consistent property of a team and was associated with the pace of the search, whereas false calls depended on the individual search and on particular items, especially hot dogs and circulated currency. Because the handler determines where, how long, and how often the dog samples, and the dog determines whether it responds, these results describe the joint work of a two-species system rather than the failing of either partner. Observing the search process in more ecologically valid conditions, rather than solely scoring its outcome, made these distinctions visible without giving up measurement.

**Acknowledgments**

We thank all the dog handlers who took part, and the research assistants, volunteers, judges, and dog trainers who supported the events. This research was supported by multiple sources. Data collection was funded by UCSD. Additional funding was provided by the Institute for Justice. The PI independently created the study design and executed implementation, analysis, and reporting. All findings, conclusions, and errors are the authors'.

## References

Amon, M. J., & Favela, L. H. (2019). Distributed cognition criteria: Defined, operationalized, and applied to human-dog systems. *Behavioural Processes, 162*, 167–176. https://doi.org/10.1016/j.beproc.2019.03.001

Bates, D., Mächler, M., Bolker, B., & Walker, S. (2015). Fitting linear mixed-effects models using lme4. *Journal of Statistical Software, 67*(1), 1–48. https://doi.org/10.18637/jss.v067.i01

Beebe, S. C., Howell, T. J., & Bennett, P. C. (2016). Using scent detection dogs in conservation settings: A review of scientific literature regarding their selection. *Frontiers in Veterinary Science, 3*, Article 96. https://doi.org/10.3389/fvets.2016.00096

Benjamini, Y., & Hochberg, Y. (1995). Controlling the false discovery rate: A practical and powerful approach to multiple testing. *Journal of the Royal Statistical Society: Series B (Methodological), 57*(1), 289–300. https://doi.org/10.1111/j.2517-6161.1995.tb02031.x

Bradski, G. (2000). The OpenCV library. *Dr. Dobb's Journal of Software Tools, 25*(11), 120–125.

Brady, K., Cracknell, N., Zulch, H., & Mills, D. S. (2018). A systematic review of the reliability and validity of behavioural tests used to assess behavioural characteristics important in working dogs. *Frontiers in Veterinary Science, 5*, Article 103. https://doi.org/10.3389/fvets.2018.00103

Bray, E. E., Otto, C. M., Udell, M. A. R., Hall, N. J., Johnston, A. M., & MacLean, E. L. (2021). Enhancing the selection and performance of working dogs. *Frontiers in Veterinary Science, 8*, Article 644431. https://doi.org/10.3389/fvets.2021.644431

Caldicott, L., Pike, T. W., Zulch, H. E., Mills, D. S., Williams, F. J., Elliker, K. R., Hutchings, B., & Wilkinson, A. (2024). Odour generalisation and detection dog training. *Animal Cognition, 27*, Article 73. https://doi.org/10.1007/s10071-024-01907-0

Chambers, J., Quinlan, M. B., Evans, A., & Quinlan, R. J. (2020). Dog-human coevolution: Cross-cultural analysis of multiple hypotheses. *Journal of Ethnobiology, 40*(4), 414–433. https://doi.org/10.2993/0278-0771-40.4.414

Clark, C. C. A., Sibbald, N. J., & Rooney, N. J. (2020). Search dog handlers show positive bias when scoring their own dog's performance. *Frontiers in Veterinary Science, 7*, Article 612. https://doi.org/10.3389/fvets.2020.00612

Cohen, J. (1960). A coefficient of agreement for nominal scales. *Educational and Psychological Measurement, 20*(1), 37–46. https://doi.org/10.1177/001316446002000104

Concha, A., Mills, D. S., Feugier, A., Zulch, H., Guest, C., Harris, R., & Pike, T. W. (2014). Using sniffing behavior to differentiate true negative from false negative responses in trained scent-detection dogs. *Chemical Senses, 39*(9), 749–754. https://doi.org/10.1093/chemse/bju045

Cox, D. R. (1972). Regression models and life-tables. *Journal of the Royal Statistical Society: Series B (Methodological), 34*(2), 187–202. https://doi.org/10.1111/j.2517-6161.1972.tb00899.x

Craven, B. A., Paterson, E. G., & Settles, G. S. (2010). The fluid dynamics of canine olfaction: Unique nasal airflow patterns as an explanation of macrosmia. *Journal of The Royal Society Interface, 7*(47), 933–943. https://doi.org/10.1098/rsif.2009.0490

DeChant, M. T., Ford, C., & Hall, N. J. (2020). Effect of handler knowledge of the detection task on canine search behavior and performance. *Frontiers in Veterinary Science, 7*, Article 250. https://doi.org/10.3389/fvets.2020.00250

Fleiss, J. L. (1971). Measuring nominal scale agreement among many raters. *Psychological Bulletin, 76*(5), 378–382. https://doi.org/10.1037/h0031619

Freedman, A. H., & Wayne, R. K. (2017). Deciphering the origin of dogs: From fossils to genomes. *Annual Review of Animal Biosciences, 5*, 281–307. https://doi.org/10.1146/annurev-animal-022114-110937

Furton, K. G., Hong, Y. C., Hsu, Y. L., Luo, T., Rose, S., & Walton, J. (2002). Identification of odor signature chemicals in cocaine using solid-phase microextraction-gas chromatography and detector dog response to isolated compounds spiked on U.S. paper currency. *Journal of Chromatographic Science, 40*(3), 147–155. https://doi.org/10.1093/chromsci/40.3.147

Gazit, I., Goldblatt, A., & Terkel, J. (2005). The role of context specificity in learning: The effects of training context on explosives detection in dogs. *Animal Cognition, 8*(3), 143–150. https://doi.org/10.1007/s10071-004-0236-9

Gibson, J. J. (1979). *The ecological approach to visual perception*. Houghton Mifflin.

Hall, N. J., Glenn, K., Smith, D. W., & Wynne, C. D. L. (2015). Performance of Pugs, German Shepherds, and Greyhounds (Canis lupus familiaris) on an odor-discrimination task. *Journal of Comparative Psychology, 129*(3), 237–246. https://doi.org/10.1037/a0039271

Hare, B., & Tomasello, M. (2005). Human-like social skills in dogs? *Trends in Cognitive Sciences, 9*(9), 439–444. https://doi.org/10.1016/j.tics.2005.07.003

Harris, C. R., Millman, K. J., van der Walt, S. J., Gommers, R., Virtanen, P., Cournapeau, D., Wieser, E., Taylor, J., Berg, S., Smith, N. J., Kern, R., Picus, M., Hoyer, S., van Kerkwijk, M. H., Brett, M., Haldane, A., del Río, J. F., Wiebe, M., Peterson, P., … Oliphant, T. E. (2020). Array programming with NumPy. *Nature, 585*(7825), 357–362. https://doi.org/10.1038/s41586-020-2649-2

Hartley, R., & Zisserman, A. (2003). *Multiple view geometry in computer vision* (2nd ed.). Cambridge University Press. https://doi.org/10.1017/CBO9780511811685

Hayes, J. E., McGreevy, P. D., Forbes, S. L., Laing, G., & Stuetz, R. M. (2018). Critical review of dog detection and the influences of physiology, training, and analytical methodologies. *Talanta, 185*, 499–512. https://doi.org/10.1016/j.talanta.2018.04.010

Hutchins, E. (1995). *Cognition in the wild*. MIT Press.

Jezierski, T., Adamkiewicz, E., Walczak, M., Sobczyńska, M., Górecka-Bruzda, A., Ensminger, J., & Papet, E. (2014). Efficacy of drug detection by fully-trained police dogs varies by breed, training level, type of drug and search environment. *Forensic Science International, 237*, 112–118. https://doi.org/10.1016/j.forsciint.2014.01.013

Jinn, J., Connor, E. G., & Jacobs, L. F. (2020). How ambient environment influences olfactory orientation in search and rescue dogs. *Chemical Senses, 45*(8), 625–634. https://doi.org/10.1093/chemse/bjaa060

Jocher, G., & Qiu, J. (2024). *Ultralytics YOLO11* (Version 11.0.0) [Computer software]. https://github.com/ultralytics/ultralytics

Johnen, D., Heuwieser, W., & Fischer-Tenhagen, C. (2017). An approach to identify bias in scent detection dog testing. *Applied Animal Behaviour Science, 189*, 1–12. https://doi.org/10.1016/j.applanim.2017.01.001

Jourdan, T. H., Veitenheimer, A. M., Murray, C. K., & Wagner, J. R. (2013). The quantitation of cocaine on U.S. currency: Survey and significance of the levels of contamination. *Journal of Forensic Sciences, 58*(3), 616–624. https://doi.org/10.1111/1556-4029.12097

Kalman, R. E. (1960). A new approach to linear filtering and prediction problems. *Journal of Basic Engineering, 82*(1), 35–45. https://doi.org/10.1115/1.3662552

Karpinsky, M., Browning, H., Quigley-McBride, A., Bunker, P., Chapman, W., Prada-Tiedemann, P. A., & DeGreeff, L. E. (2025). Explosive detection canines in the field: A multi-site black box validation study. *Frontiers in Veterinary Science, 12*, Article 1668317. https://doi.org/10.3389/fvets.2025.1668317

Kaufhold, S. P., Borzello, M., Rossano, F., & Kirsh, D. (2026). Beyond the lab: Cognitive neuroscience in real-world contexts. *WIREs Cognitive Science, 17*(2), Article e70027. https://doi.org/10.1002/wcs.70027

Kuznetsova, A., Brockhoff, P. B., & Christensen, R. H. B. (2017). lmerTest package: Tests in linear mixed effects models. *Journal of Statistical Software, 82*(13), 1–26. https://doi.org/10.18637/jss.v082.i13

La Toya, J. J., Baxter, G. S., & Murray, P. J. (2017). Identifying suitable detection dogs. *Applied Animal Behaviour Science, 195*, 1–7. https://doi.org/10.1016/j.applanim.2017.06.010

Lazarowski, L., & Dorman, D. C. (2014). Explosives detection by military working dogs: Olfactory generalization from components to mixtures. *Applied Animal Behaviour Science, 151*, 84–93. https://doi.org/10.1016/j.applanim.2013.11.010

Lazarowski, L., Krichbaum, S., DeGreeff, L. E., Simon, A., Singletary, M., Angle, C., & Waggoner, L. P. (2020a). Methodological considerations in canine olfactory detection research. *Frontiers in Veterinary Science, 7*, Article 408. https://doi.org/10.3389/fvets.2020.00408

Lazarowski, L., Waggoner, L. P., Krichbaum, S., Singletary, M., Haney, P., Rogers, B., & Angle, C. (2020b). Selecting dogs for explosives detection: Behavioral characteristics. *Frontiers in Veterinary Science, 7*, Article 597. https://doi.org/10.3389/fvets.2020.00597

Lazarowski, L., Rogers, B., Smith, J. G., Krichbaum, S., & Waggoner, P. (2023). Longitudinal stability of detection dog behavioral assessment: A follow-up study of long-term working success. *Applied Animal Behaviour Science, 268*, Article 106082. https://doi.org/10.1016/j.applanim.2023.106082

Lit, L., Schweitzer, J. B., & Oberbauer, A. M. (2011). Handler beliefs affect scent detection dog outcomes. *Animal Cognition, 14*(3), 387–394. https://doi.org/10.1007/s10071-010-0373-2

Martin, C., Willem, N., Desablens, S., Menard, V., Tajri, S., Blanchard, S., Brostaux, Y., Verheggen, F., & Diederich, C. (2022). What a good boy! Deciphering the efficiency of detection dogs. *Frontiers in Analytical Science, 2*, Article 932857. https://doi.org/10.3389/frans.2022.932857

Martvel, G., Pedretti, G., Lazebnik, T., Zamansky, A., Ouchi, Y., Monteiro, T., Farhat, N., Shimshoni, I., Grinstein, D., Michaeli, Y., Valsecchi, P., Hall, N., & Marshall-Pescini, S. (2025). Does the tail show when the nose knows? Artificial intelligence outperforms human experts at predicting detection dogs finding their target through tail kinematics. *Royal Society Open Science, 12*(8), Article 250399. https://doi.org/10.1098/rsos.250399

Mathis, A., Mamidanna, P., Cury, K. M., Abe, T., Murthy, V. N., Mathis, M. W., & Bethge, M. (2018). DeepLabCut: Markerless pose estimation of user-defined body parts with deep

learning. *Nature Neuroscience, 21*(9), 1281–1289. https://doi.org/10.1038/s41593-018-0209-y

McKinney, W. (2010). Data structures for statistical computing in Python. In S. van der Walt & J. Millman (Eds.), *Proceedings of the 9th Python in Science Conference* (pp. 56–61). https://doi.org/10.25080/Majora-92bf1922-00a

Mejia, D., Burnett, L., Hebdon, N., Stevens, P., Shiber, A., Cranston, C., DeGreeff, L., & Waldrop, L. D. (2024). Physical properties of odorants affect behavior of trained detection dogs during close-quarters searches. *Scientific Reports, 14*, Article 4843. https://doi.org/10.1038/s41598-024-55323-y

Miklósi, Á., & Topál, J. (2013). What does it take to become "best friends"? Evolutionary changes in canine social competence. *Trends in Cognitive Sciences, 17*(6), 287–294. https://doi.org/10.1016/j.tics.2013.04.005

Naderi, Sz., Miklósi, Á., Dóka, A., & Csányi, V. (2001). Co-operative interactions between blind persons and their dogs. *Applied Animal Behaviour Science, 74*(1), 59–80. https://doi.org/10.1016/S0168-1591(01)00152-6

Nath, T., Mathis, A., Chen, A. C., Patel, A., Bethge, M., & Mathis, M. W. (2019). Using DeepLabCut for 3D markerless pose estimation across species and behaviors. *Nature Protocols, 14*(7), 2152–2176. https://doi.org/10.1038/s41596-019-0176-0

Paszke, A., Gross, S., Massa, F., Lerer, A., Bradbury, J., Chanan, G., Killeen, T., Lin, Z., Gimelshein, N., Antiga, L., Desmaison, A., Köpf, A., Yang, E., DeVito, Z., Raison, M., Tejani, A., Chilamkurthy, S., Steiner, B., Fang, L., … Chintala, S. (2019). PyTorch: An imperative style, high-performance deep learning library. In *Advances in Neural Information Processing Systems 32* (pp. 8024–8035). Curran Associates.

Pedersen, T. L. (2024). *patchwork: The composer of plots* (R package version 1.2.0). https://CRAN.R-project.org/package=patchwork

Pfungst, O. (1911). *Clever Hans (the horse of Mr. von Osten): A contribution to experimental animal and human psychology* (C. L. Rahn, Trans.). Henry Holt.

Porritt, F., Shapiro, M., Waggoner, P., Mitchell, E., Thomson, T., Nicklin, S., & Kacelnik, A. (2015). Performance decline by search dogs in repetitive tasks, and mitigation strategies. *Applied Animal Behaviour Science, 166*, 112–122. https://doi.org/10.1016/j.applanim.2015.02.013

Poupko, J. M., Hearn, W. L., & Rossano, F. (2018). Drug contamination of U.S. paper currency and forensic relevance of canine alert to paper currency: A critical review of the scientific literature. *Journal of Forensic Sciences, 63*(5), 1340–1345. https://doi.org/10.1111/1556-4029.13755

Quigley-McBride, A., Prada-Tiedemann, P. A., & Helfers, F. (2026). Strengthening operational performance in canine detection teams with double-blind certification testing. *Forensic Science International: Synergy, 12*, Article 100658. https://doi.org/10.1016/j.fsisyn.2026.100658

R Core Team. (2024). *R: A language and environment for statistical computing* (Version 4.3.3). R Foundation for Statistical Computing. https://www.R-project.org/

Rauch, H. E., Tung, F., & Striebel, C. T. (1965). Maximum likelihood estimates of linear dynamic systems. *AIAA Journal, 3*(8), 1445–1450. https://doi.org/10.2514/3.3166

Rice, B. L., & Velasco, J. G. (2021). Accuracy comparison of single-purpose and dual-purpose narcotic detection canines. *Applied Animal Behaviour Science, 243*, Article 105434. https://doi.org/10.1016/j.applanim.2021.105434

Rice, B. L., & Velasco, J. (2023). Comparing narcotics detection canine accuracy across breeds. *Heliyon, 9*(8), Article e19040. https://doi.org/10.1016/j.heliyon.2023.e19040

Rooney, N. J., & Clark, C. C. A. (2021). Development of a performance monitoring instrument for rating explosives search dog performance. *Frontiers in Veterinary Science, 8*, Article 545382. https://doi.org/10.3389/fvets.2021.545382

Roulaux, P. E. M., Vinke, C. M., van Herwijnen, I. R., & Arndt, S. S. (2026). Exploring ideal scent detection dog characteristics for successful professional deployment as derived from face-to-face structured interviews with professional scent detection dog handlers. *PLOS ONE, 21*(2), Article e0339379. https://doi.org/10.1371/journal.pone.0339379

Rutter, N. J., Howell, T. J., Stukas, A. A., Pascoe, J. H., & Bennett, P. C. (2021). Diving in nose first: The influence of unfamiliar search scale and environmental context on the search performance of volunteer conservation detection dog–handler teams. *Animals, 11*(4), Article 1177. https://doi.org/10.3390/ani11041177

Salomons, H., Smith, K. C. M., Callahan-Beckel, M., Callahan, M., Levy, K., Kennedy, B. S., Bray, E. E., Gnanadesikan, G. E., Horschler, D. J., Gruen, M., Tan, J., White, P., vonHoldt, B. M., MacLean, E. L., & Hare, B. (2021). Cooperative communication with humans evolved to emerge early in domestic dogs. *Current Biology, 31*(14), 3137–3144.e11. https://doi.org/10.1016/j.cub.2021.06.051

Shrout, P. E., & Fleiss, J. L. (1979). Intraclass correlations: Uses in assessing rater reliability. *Psychological Bulletin, 86*(2), 420–428. https://doi.org/10.1037/0033-2909.86.2.420

Slowikowski, K., van den Brand, T., Schep, A., Hughes, S., Dang, T. K., Lukauskas, S., Irisson, J. O., Kamvar., Z. N., Ryan, T., Christophe, D., Hiroaki, Y., Gramme, P., Abdol, A. M., Barrett, M., Cannoodt, R., Krassowski, M., Chirico, M., Aphalo, P., Barton, F. (2024).

*ggrepel: Automatically position non-overlapping text labels with 'ggplot2'* (R package version 0.9.5). https://CRAN.R-project.org/package=ggrepel

Therneau, T. (2024). *A package for survival analysis in R* (R package version 3.5-8). https://CRAN.R-project.org/package=survival

Tiira, K., Tikkanen, A., & Vainio, O. (2020). Inhibitory control – Important trait for explosive detection performance in police dogs? *Applied Animal Behaviour Science, 224*, Article 104942. https://doi.org/10.1016/j.applanim.2020.104942

Triggs, B., McLauchlan, P. F., Hartley, R. I., & Fitzgibbon, A. W. (2000). Bundle adjustment: A modern synthesis. In B. Triggs, A. Zisserman, & R. Szeliski (Eds.), *Vision algorithms: Theory and practice* (pp. 298–372). Springer. https://doi.org/10.1007/3-540-44480-7_21

Troisi, C. A., Mills, D. S., Wilkinson, A., & Zulch, H. E. (2019). Behavioral and cognitive factors that affect the success of scent detection dogs. *Comparative Cognition & Behavior Reviews, 14*, 51–76. https://doi.org/10.3819/CCBR.2019.140007

Uexküll, J. von. (2010). *A foray into the worlds of animals and humans: With A theory of meaning* (J. D. O'Neil, Trans.). University of Minnesota Press. (Original work published 1934)

Virtanen, P., Gommers, R., Oliphant, T. E., Haberland, M., Reddy, T., Cournapeau, D., Burovski, E., Peterson, P., Weckesser, W., Bright, J., van der Walt, S. J., Brett, M., Wilson, J., Millman, K. J., Mayorov, N., Nelson, A. R. J., Jones, E., Kern, R., Larson, E., … van Mulbregt, P. (2020). SciPy 1.0: Fundamental algorithms for scientific computing in Python. *Nature Methods, 17*(3), 261–272. https://doi.org/10.1038/s41592-019-0686-2

Wickham, H. (2016). *ggplot2: Elegant graphics for data analysis*. Springer. https://ggplot2.tidyverse.org

Wittenburg, P., Brugman, H., Russel, A., Klassmann, A., & Sloetjes, H. (2006). ELAN: A professional framework for multimodality research. In *Proceedings of the Fifth International Conference on Language Resources and Evaluation (LREC'06)* (pp. 1556–1559). European Language Resources Association.

Ye, S., Filippova, A., Lauer, J., Schneider, S., Vidal, M., Qiu, T., Mathis, A., & Mathis, M. W. (2024). SuperAnimal pretrained pose estimation models for behavioral analysis. *Nature Communications, 15*, Article 5165. https://doi.org/10.1038/s41467-024-48792-2

Zeileis, A., & Hothorn, T. (2002). Diagnostic checking in regression relationships. *R News, 2*(3), 7–10. https://CRAN.R-project.org/doc/Rnews/

Zeileis, A., Köll, S., & Graham, N. (2020). Various versatile variances: An object-oriented implementation of clustered covariances in R. *Journal of Statistical Software, 95*(1), 1–36. https://doi.org/10.18637/jss.v095.i01

Zhang, Z. (2000). A flexible new technique for camera calibration. *IEEE Transactions on Pattern Analysis and Machine Intelligence, 22*(11), 1330–1334. https://doi.org/10.1109/34.888718

Zheng, J., Lazarowski, L., Lanier, A. L., Haney, P. S., Chester, E. M., Waggoner, P., & Wang, X. (2026). Behavioral components define operational suitability metric for detection dog success. *Frontiers in Veterinary Science, 13*, Article 1887925. https://doi.org/10.3389/fvets.2026.1887925

Zubedat, S., Aga-Mizrachi, S., Cymerblit-Sabba, A., Shwartz, J., Leon, J. F., Rozen, S., Varkovitzky, I., Eshed, Y., Grinstein, D., & Avital, A. (2014). Human–animal interface: The effects of handler's stress on the performance of canines in an explosive detection task. *Applied Animal Behaviour Science, 158*, 69–75.

https://doi.org/10.1016/j.applanim.2014.05.004